\documentclass[aps,prb,notitlepage,amssymb,floats,superscriptaddress]{revtex4-2}

\usepackage{graphicx}   
\usepackage{amsmath}
\usepackage{epstopdf}
\usepackage{listings}
\usepackage{mathtools}
\usepackage{hyperref}

\begin{document}
	\title{Paring density waves as the origin of a ring-like RIXS profile in Bi$_2$Sr$_2$CaCu$_2$O$_{8+\delta}$}
	\author{David Dentelski}
	\affiliation{Department of Physics, Bar-Ilan University, Ramat Gan 5290002, Israel}
	\affiliation{Center for Quantum Entanglement Science and Technology, Bar-Ilan University, Ramat Gan 5290002, Israel}
	\author{Emanuele G. Dalla Torre}
	\affiliation{Department of Physics, Bar-Ilan University, Ramat Gan 5290002, Israel}
	\affiliation{Center for Quantum Entanglement Science and Technology, Bar-Ilan University, Ramat Gan 5290002, Israel}
	\date{\today}

	\maketitle

	{\bf The coexistence of a homogeneous d-wave gap and short-ranged pairing density waves (PDW) accounts for the apparent ``ring" charge order in all directions of the copper oxide plane, observed by recent RIXS measurements in Bi$_2$Sr$_2$CaCu$_2$O$_{8+\delta}$ (Bi2212) \cite{boschini2021dynamic}}.
	\\ \\
	Almost twenty years ago, scanning tunneling experiments found incommensurate density waves on the surface of Bi$_2$Sr$_2$CaCu$_2$O$_{8+\delta}$ (BSCCO) \cite{hoffman2002four, howald2003periodic, vershinin2004local, hanaguri2004checkerboard}. Ten years later, resonant X-ray scattering experiments detected a similar incommensurate order in the bulk of YBa$_2$Cu$_3$O$_{7-\delta}$ (YBCO) \cite{ghiringhelli2012long}. The same order was later found in a large number of cuprates, demonstrating that this effect is ubiquitous \cite{chang2012direct,torchinsky2013fluctuating,blackburn2013x,comin2014charge,da2014ubiquitous,le2014inelastic,hashimoto2014direct,tabis2014charge,huecker2014competing,achkar2014impact,gerber2015three,hamidian2015magnetic,peng2016direct, chaix2017dispersive, peng2018re,  jang2017superconductivity,da2018coupling,bluschke2019adiabatic, kang2019evolution}.  A common approach claims that these modulations are due to a charge density wave (CDW) order that competes with superconductivity. In a recent publication \cite{dentelski2020minimal}, we claimed that short-ranged pairing density waves (PDW) within a d-wave superconducting phase are the correct interpenetration of the experimental results. Using a weak-coupling approach, we demonstrated that PDWs and CDWs lead to distinctive momentum and energy dependencies: for PDW the scattering signal peaks at momentum $(\pm q,0) , (0, \pm q)$, while for CDW it peaks at $(\pm q,\pm q)$. In addition, the PDW signal is mainly elastic $\Omega = 0$, in contrast to the CDW case, where the signal is peaked at $\Omega = \pm 2\Delta_{0}$. The experimentally detected signal is peaked at $(\pm q,0)$, $(0, \pm q)$ and is visible at $\Omega=0$, indicating that the observed density waves have a predominant PDW nature. Interestingly, the predicted scattering intensity had weaker peaks in the $(\pm q,\pm q)$ direction, which we interpreted as weak CDW modulations, born from the interplay between the static d-wave order and the short ranged PDW.

	\begin{figure}  [h]
		\centering
		\begin{tabular}{c c c}
			(a) p = 0.1  & (b) p = 0.2 & (c) p = 0.25\\
			\includegraphics[width=0.31\textwidth]{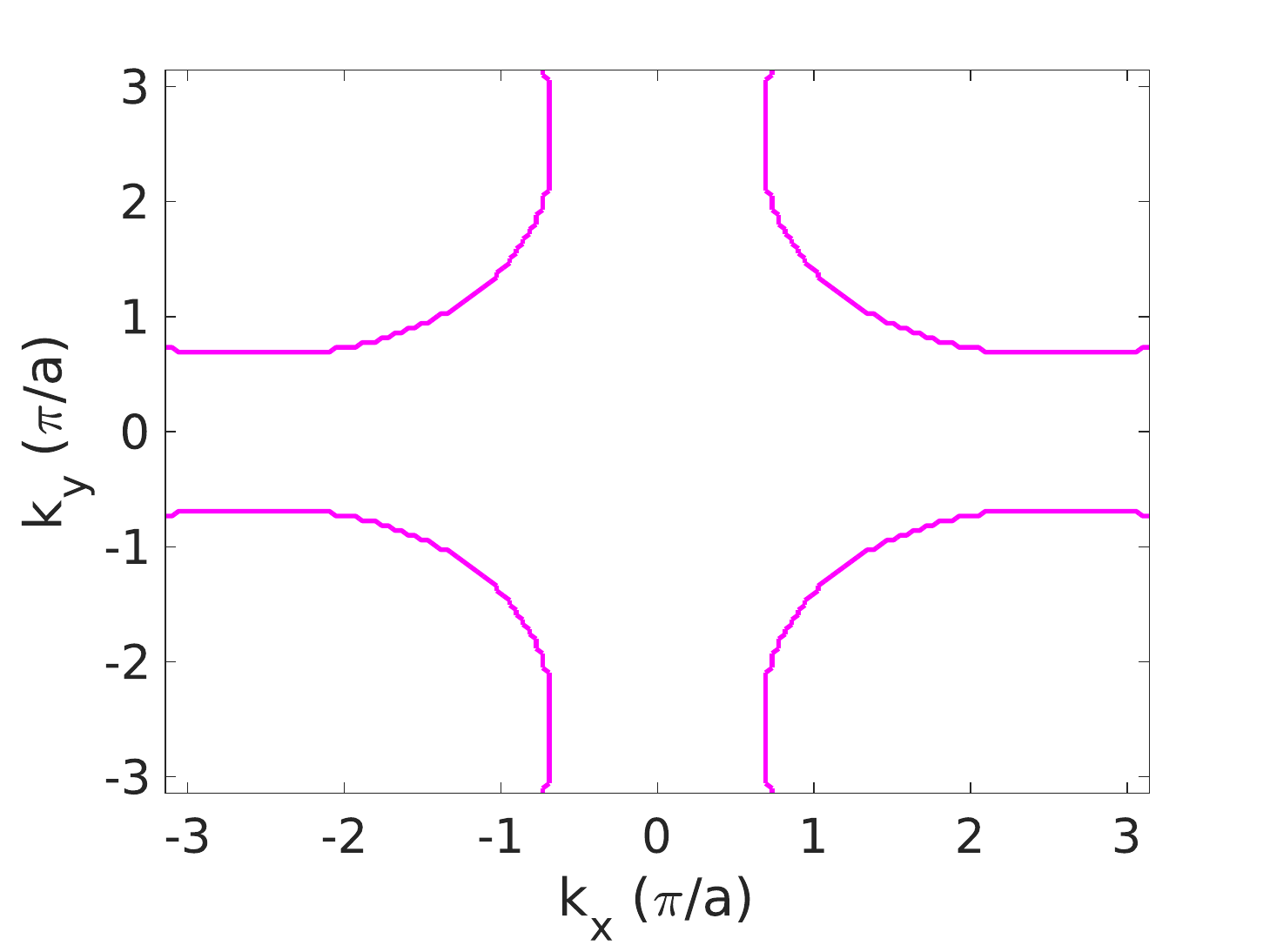}&
			\includegraphics[width=0.31\textwidth]{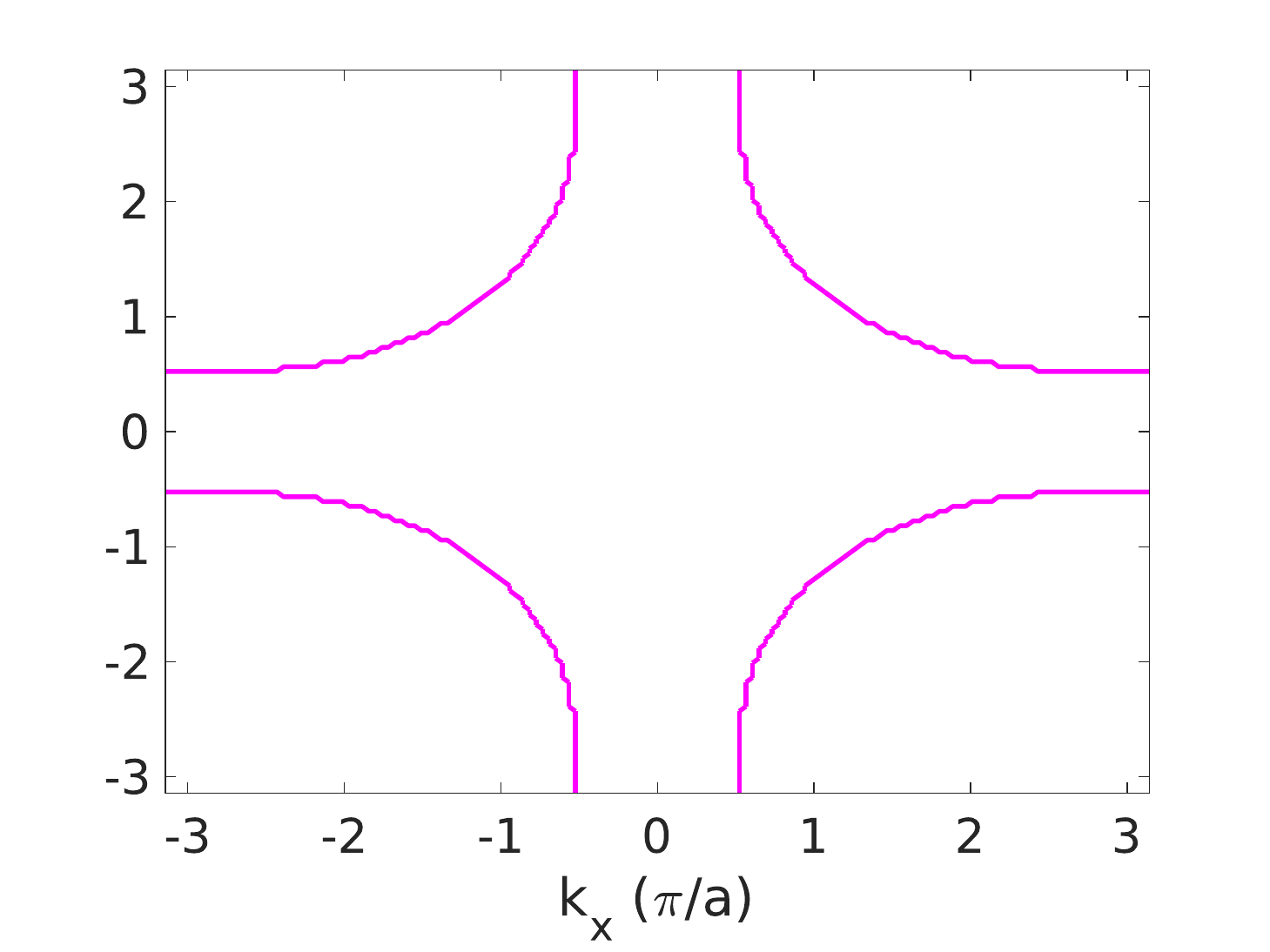}&
			\includegraphics[width=0.31\textwidth]{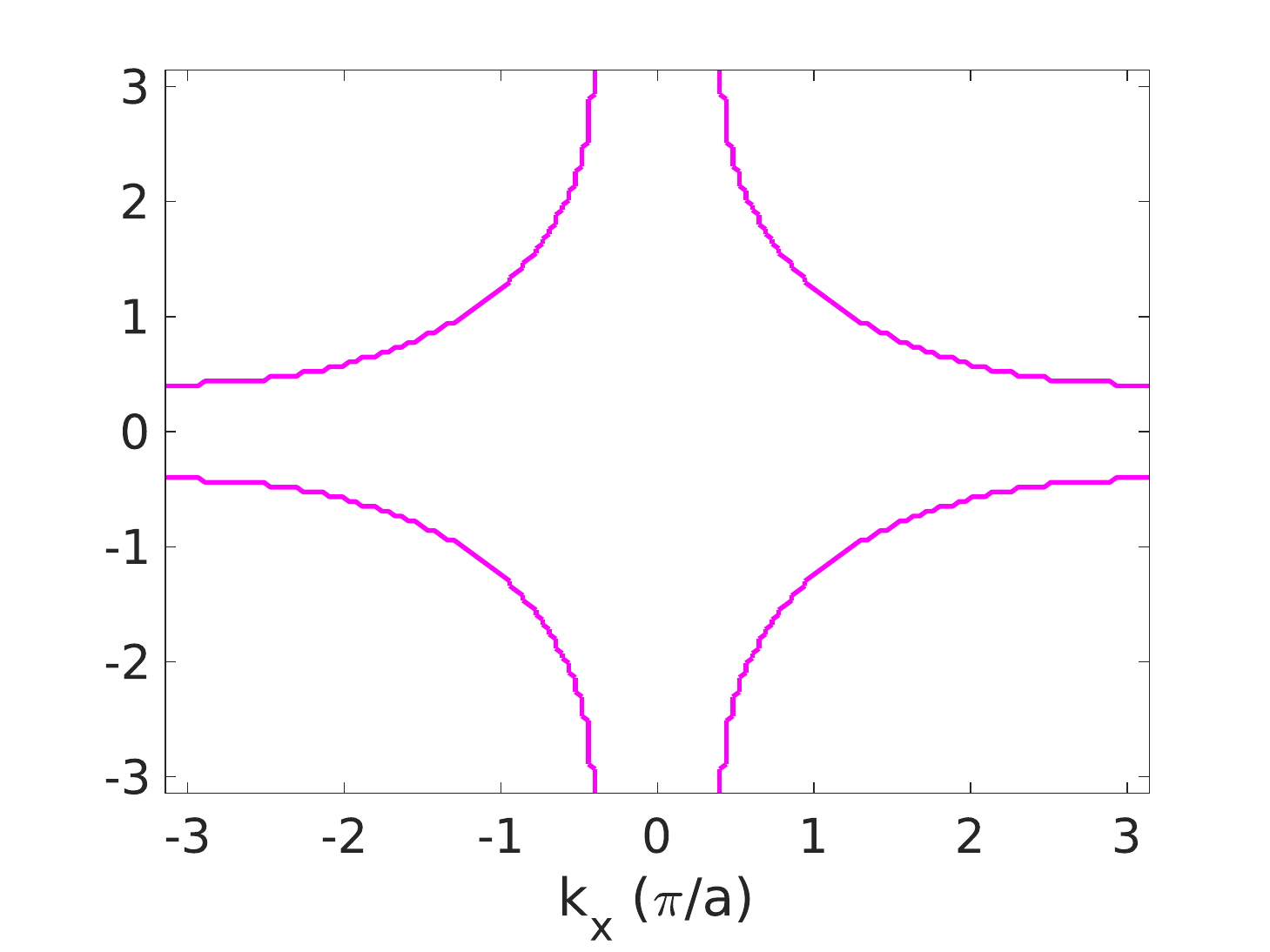}\\
			\includegraphics[width=0.31\textwidth]{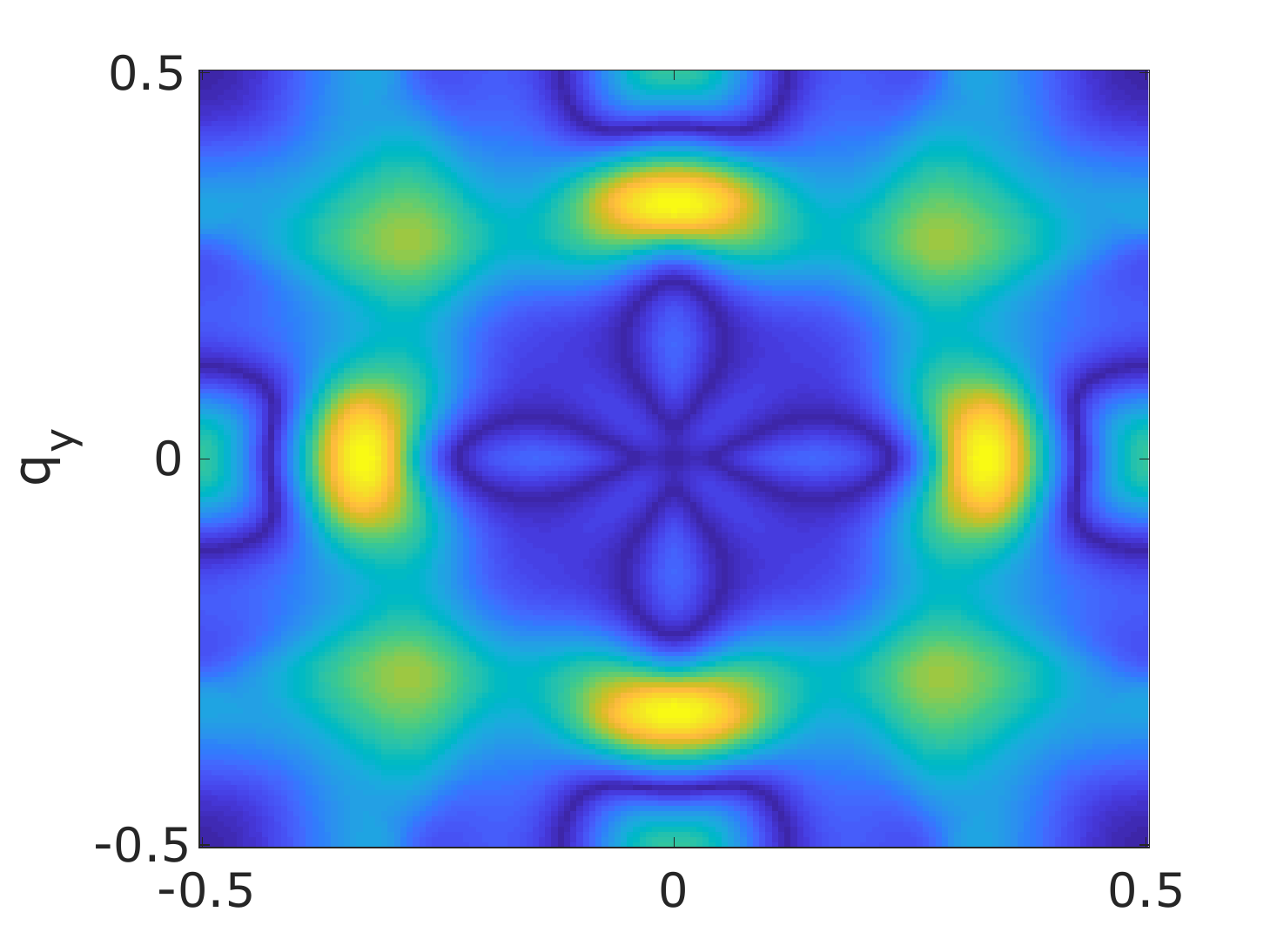}&
			\includegraphics[width=0.31\textwidth]{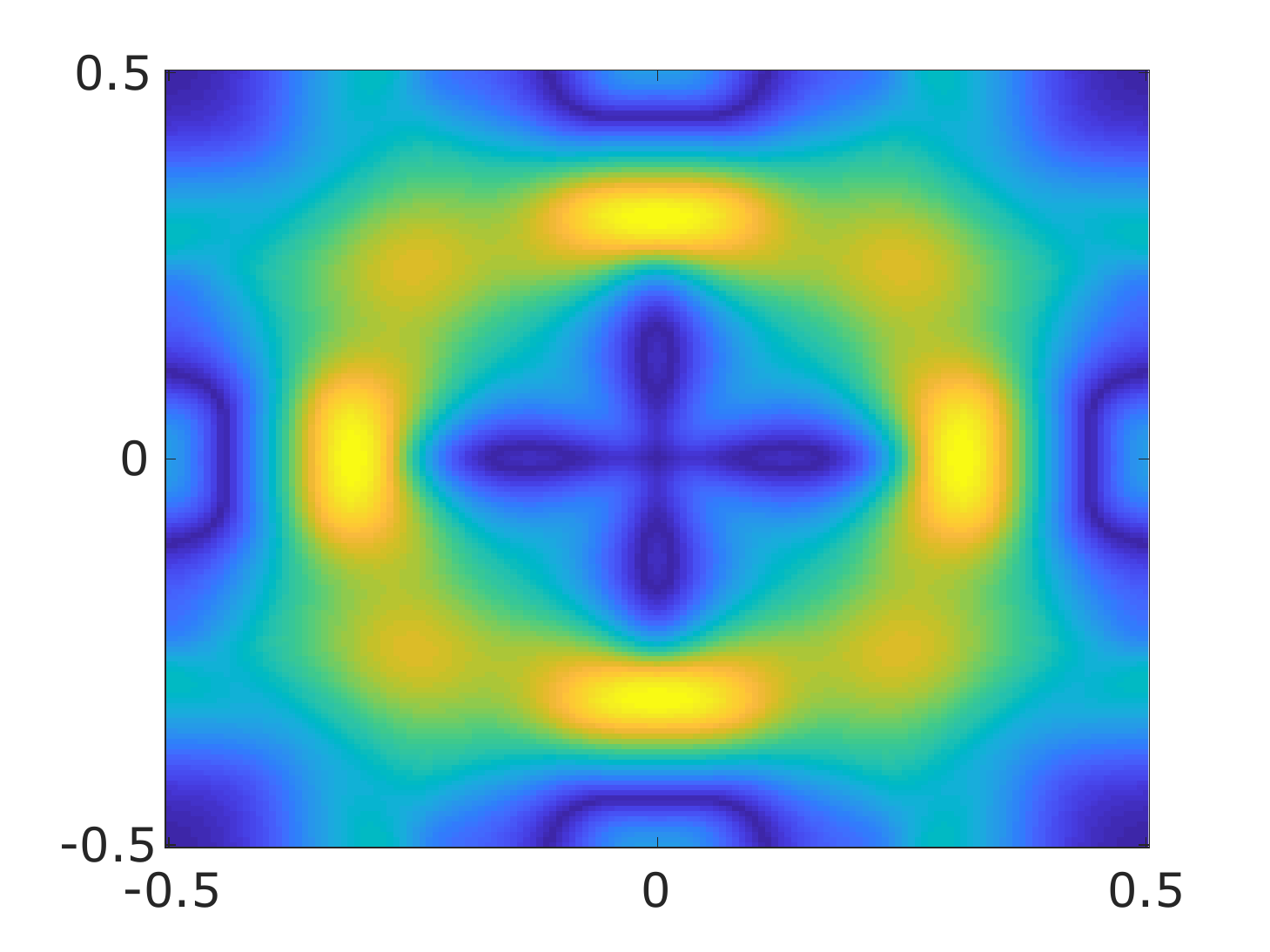}&
			\includegraphics[width=0.31\textwidth]{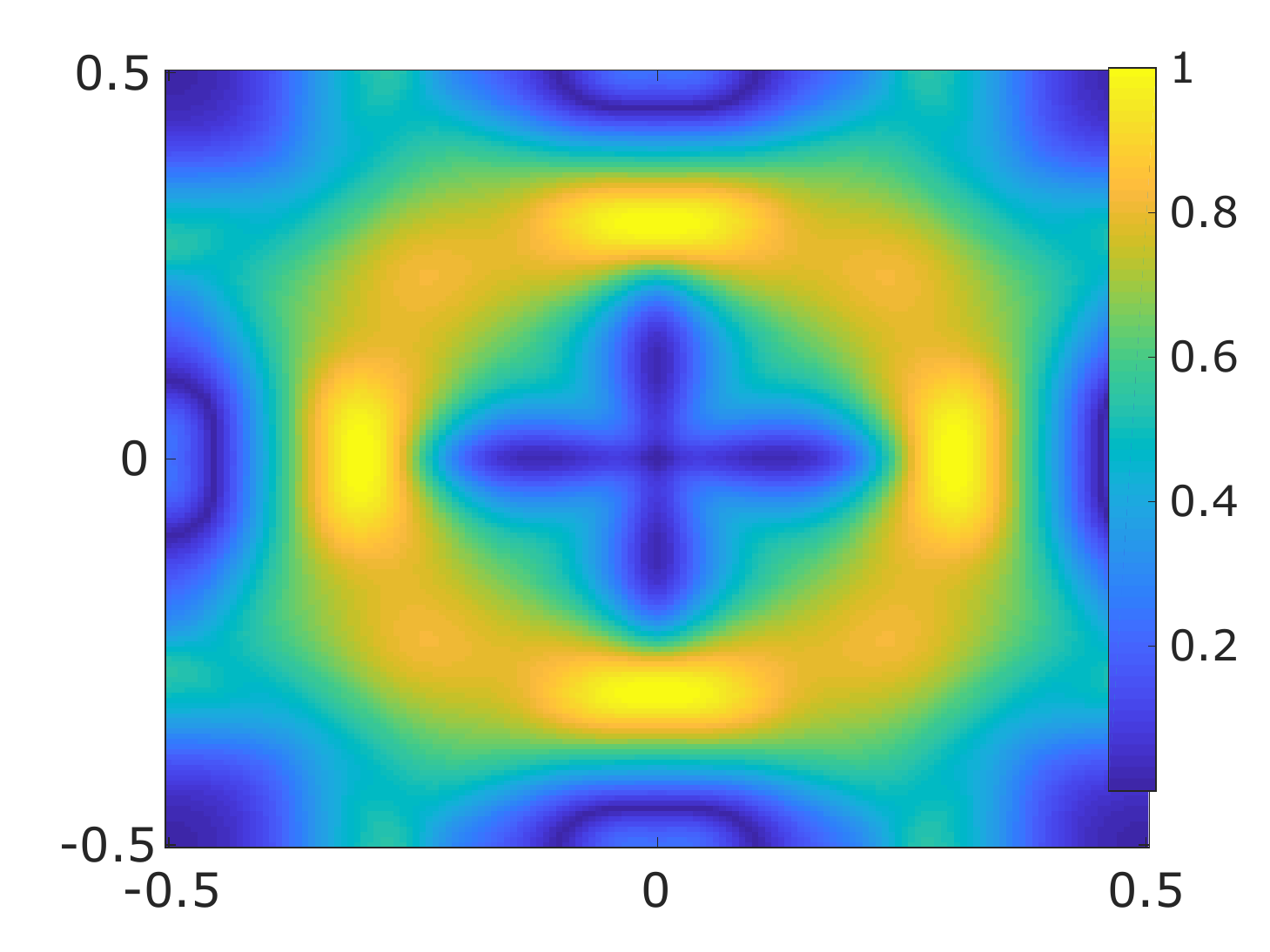}\\
			\includegraphics[width=0.31\textwidth]{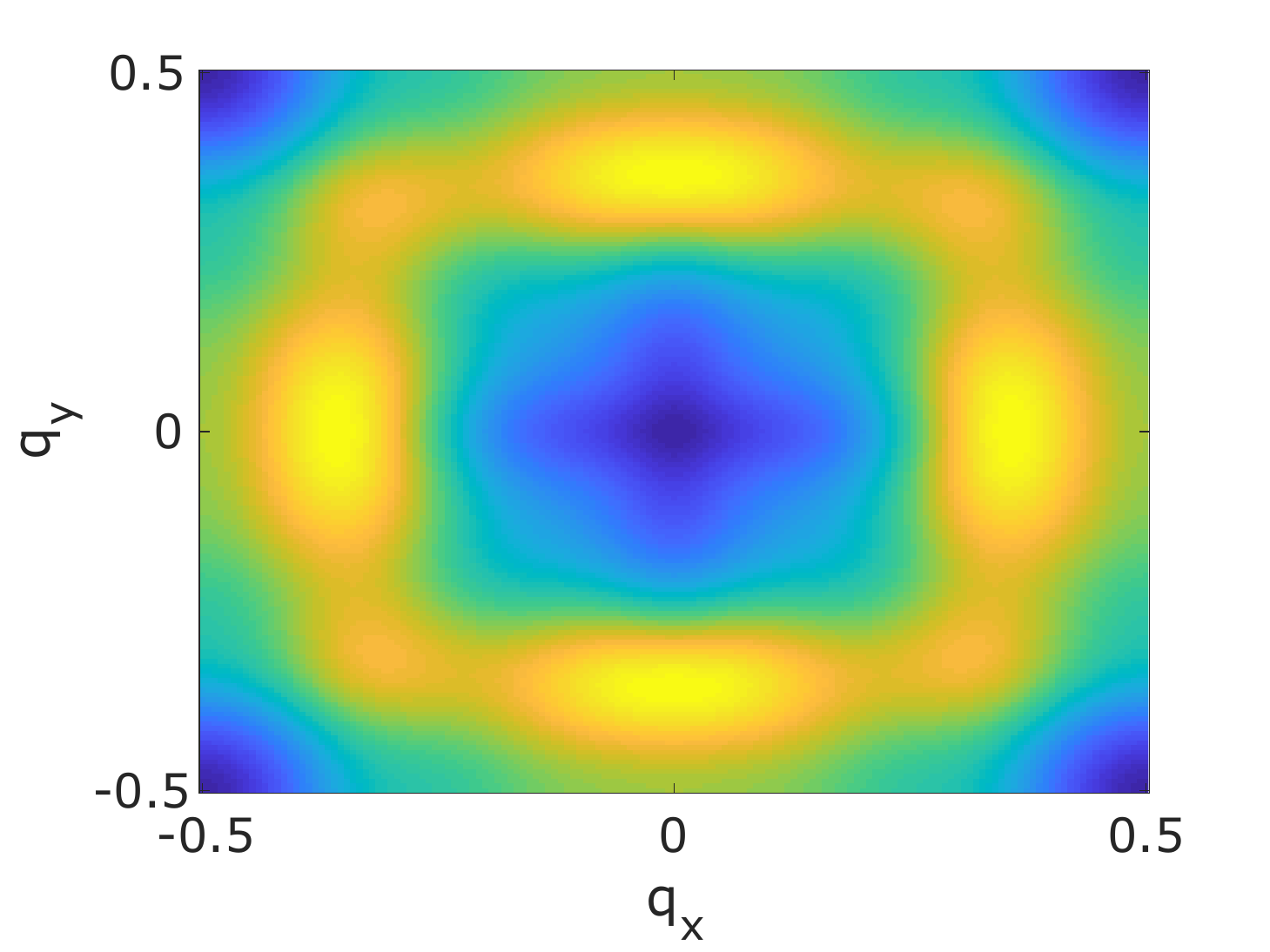}&
			\includegraphics[width=0.31\textwidth]{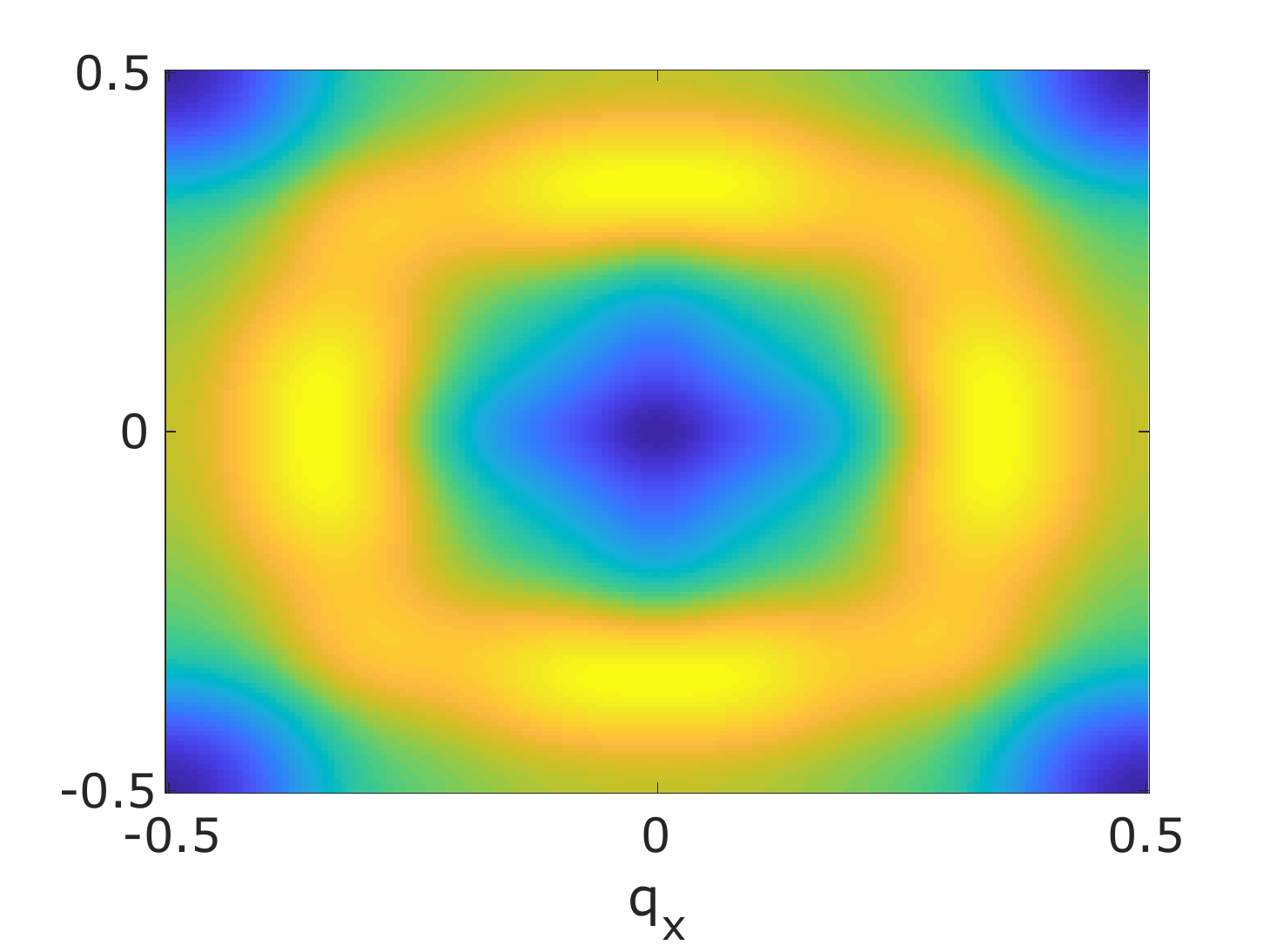}&
			\includegraphics[width=0.31\textwidth]{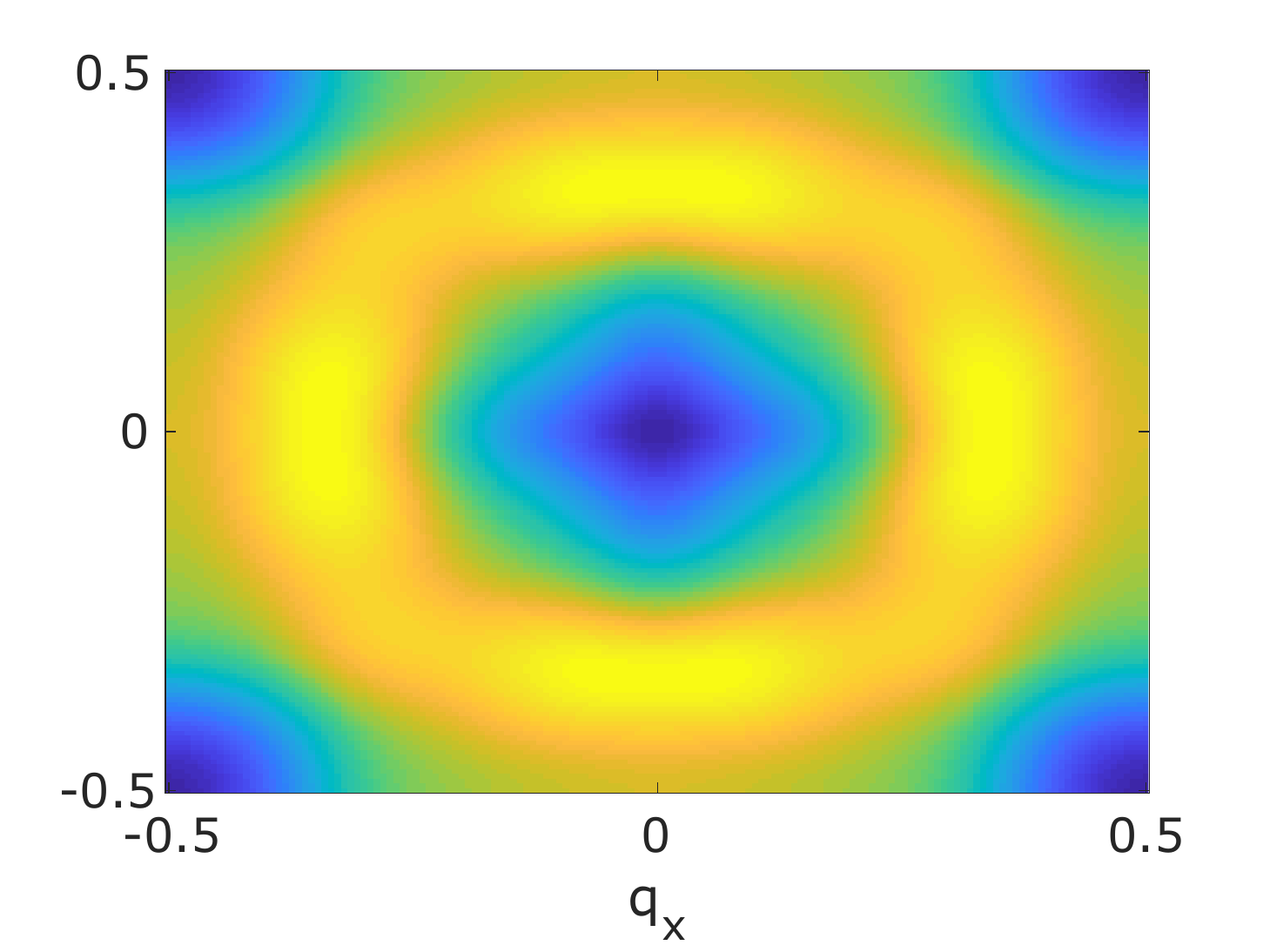}
		\end{tabular}
		\caption {Full 2D maps of the density response, Eq.~(\ref{eq:rhoq3}) normalized by its maximum, for different doping and energy values. The upper panel shows the Fermi surfaces for each doping, and the middle one is the corresponding elastic component ($\Omega =0$) of the intensity. As the doping increases, the peak move to a smaller momentum and the signal acquires a ring shape. In the lower panel we present the same maps but for finite energy $\Omega = 0.2$eV, which results in broader peaks for all doping levels.   }
		\label{Main}
	\end{figure}

	Recently, a state-of-the-art experiment was performed using resonant inelastic x-ray scattering (RIXS), spanning the entire copper-oxide plane of Bi2212 \cite{boschini2021dynamic}. The main result was the existence of a ring-like intensity peak in the $q_x-q_y$ plane with radius of $q\approx 0.27$ rlu, but with stronger signal along $(\pm q,0)$ and $(0,\pm q)$. The authors explained this signal by considering the Coulomb interaction between valance electrons, with short and long-range contributions, although concluding that this explanation is insufficient to account for the full intensity profile. In addition, motivated by the fact that the quasi-elastic ($-200<E<200$meV) intensity is more strongly peaked at $(\pm q,0), (0,\pm q)$ with respect to the higher energy signal ($500<E<900$meV), they suggested a scenario of static directional CDW, combined with dynamic fluctuating ones. To account for these fluctuations, the authors considered the effect of the dynamic susceptibility,  characterized by the Lindhard function. However, as they showed in the Supplementary Material (Note 5), this mainly yields peaks in the wrong direction - $(\pm q,\pm q)$ -  and has a square-like shape which maintains a $C_4$ symmetry, unlike the smooth ring shape observed in the experiment.

	Although we agree that this form of the Lindhard function (which was first suggested by us as an explanation for resonant x-ray experiments in Ref.~\cite{dalla2016friedel}) fails to provide the correct profile of the intensity, we believe that the authors' choice to discard Fermi-surface effects is incorrect. Indeed, if the observed signal is due to PDW fluctuations, rather than CDW ones, one needs to compute the response from a pairing-like impurity in the presence of a constant d-wave pairing gap \cite{dentelski2020minimal}. The Born approximation leads to the density response
	\begin{align} \label{eq:rhoq}
		\chi(\textbf{q},\Omega) = \int d\omega \int d^{d}k~\rm Tr \left[G_{0}(\textbf{k},\omega) V_{\bf k} G_{0}(\textbf{k}+\textbf{q}, \omega+\Omega)\sigma^z\right].
	\end{align}
	Here, $\rm V_\textbf{k} = \Delta_{\textbf{k}}\sigma^{x}$ models the impurity, $\sigma^j, j=x,z$ are Pauli matrices, $\Delta_{\textbf{k}}=\dfrac{\Delta_0}{2} (\cos(k_{x})-\cos(k_{y}))$ is the pairing gap and  $\rm G_0$ is the bare Green's function $
	\rm G^{-1}_{0}({\bf k},\omega)=
	\begin{psmallmatrix} - \omega +\varepsilon_{\textbf{k}}-\mu& \Delta_{\textbf{k}} \\  \Delta^{\star}_{\textbf{k}} &-\omega-\varepsilon_{\textbf{k}}+\mu\end{psmallmatrix},$
	where $\varepsilon_{\textbf{k}}$ is the band structure of the material, and $\mu$ the chemical potential. By performing the integral over $\omega$, one obtains
	\begin{align}\label{eq:rhoq3}
		\chi(\textbf{q},\Omega) =2\pi \int d^2 k~ \Delta_{\bf k}\frac{\varepsilon_{\bf k}\Delta_{\bf k+q}+\varepsilon_{\bf k+q}\Delta_{\bf k}}{(E_{\bf k}-E_{\bf k+q})^2+(\Omega-i
			\Gamma)^2}\left(\dfrac1{E_\textbf{k}}-\dfrac1{E_{\textbf{k}+\textbf{q}}}\right),
	\end{align}
	where $E_{\textbf{k}} = \sqrt{\varepsilon^{2}_\textbf{k}+\Delta_{\textbf{k}}^2}$ and $\Gamma$ is set by the maximum between the quasiparticles' inverse lifetime and the experimental energy resolution.
	This approach leads to a strong peak in the $(\pm q,0), (0, \pm q)$ direction, along with weaker maxima in all directions (see Fig.~2 of Ref.~\cite{dentelski2020minimal}). In Ref.~\cite{dentelski2020minimal}, we considered a minimal model of the Fermi surface, where the band structure includes only nearest- and next-nearest-neighbor hopping, allowing us to highlight the generality of our results and to validate them by a real-space extended Hubbard model.

	In this comment, we show that the experimental signal of Ref.~\cite{boschini2021dynamic} can be  reproduced by considering a realistic Fermi surface that includes longer-range couplings. Specifically, we use here the phenomenological band structure of Bi2212 proposed by Ref.~\cite{BandStructure}, that is: $\varepsilon_{k} = 0.5t_{0} (\cos(k_x)+\cos(k_y))+t_{1}\cos(k_x) \cos(k_y)+0.5t_{2}(\cos(2k_x)+\cos(2k_y))+0.5t_{3}(\cos(2k_x) \cos(k_y)+\cos(2k_y) \cos(k_x))+t_{4}\cos(2k_x) \cos(2k_y)$, with $t_{0} = -0.5951, t_{1} = 0.1636, t_{2} =-0.0519, t_{3} = -0.1117, t_{4} = 0.510$eV. The chemical potential $\mu$ is fixed by the doping through Luttinger count, and the only free parameters are $\Delta_{0}$ and $\Gamma$, which we set to the experimentally relevant values of $\Delta_{0}=0.3$ and $\Gamma=0.05$eV.

	Our main results are presented in Fig.~\ref{Main}, where the two top panels show, respectively, the phenomenological Fermi surfaces for different doping levels and the elastic component ($\Omega = 0$) of the predicted RIXS signal, Eq.~(\ref{eq:rhoq3}). At small doping ($p = 0.1$), the pronounced PDW peaks at $(\pm q,0), (0, \pm q)$ are accompanied by weaker peaks at $(\pm q, \pm q)$, associated with secondary CDW modulations. As the doping increases, the nesting areas in the Fermi surface become less parallel, leading to a broadening of the signal. As a result, the intensity profile shifts from square to a ring-like shape.

\begin{figure}[h]
		\begin{tabular}{c c c}
			(a) p = 0.1  & (b) p = 0.2 & (c) p = 0.25\\
			\includegraphics[width=0.32\textwidth]{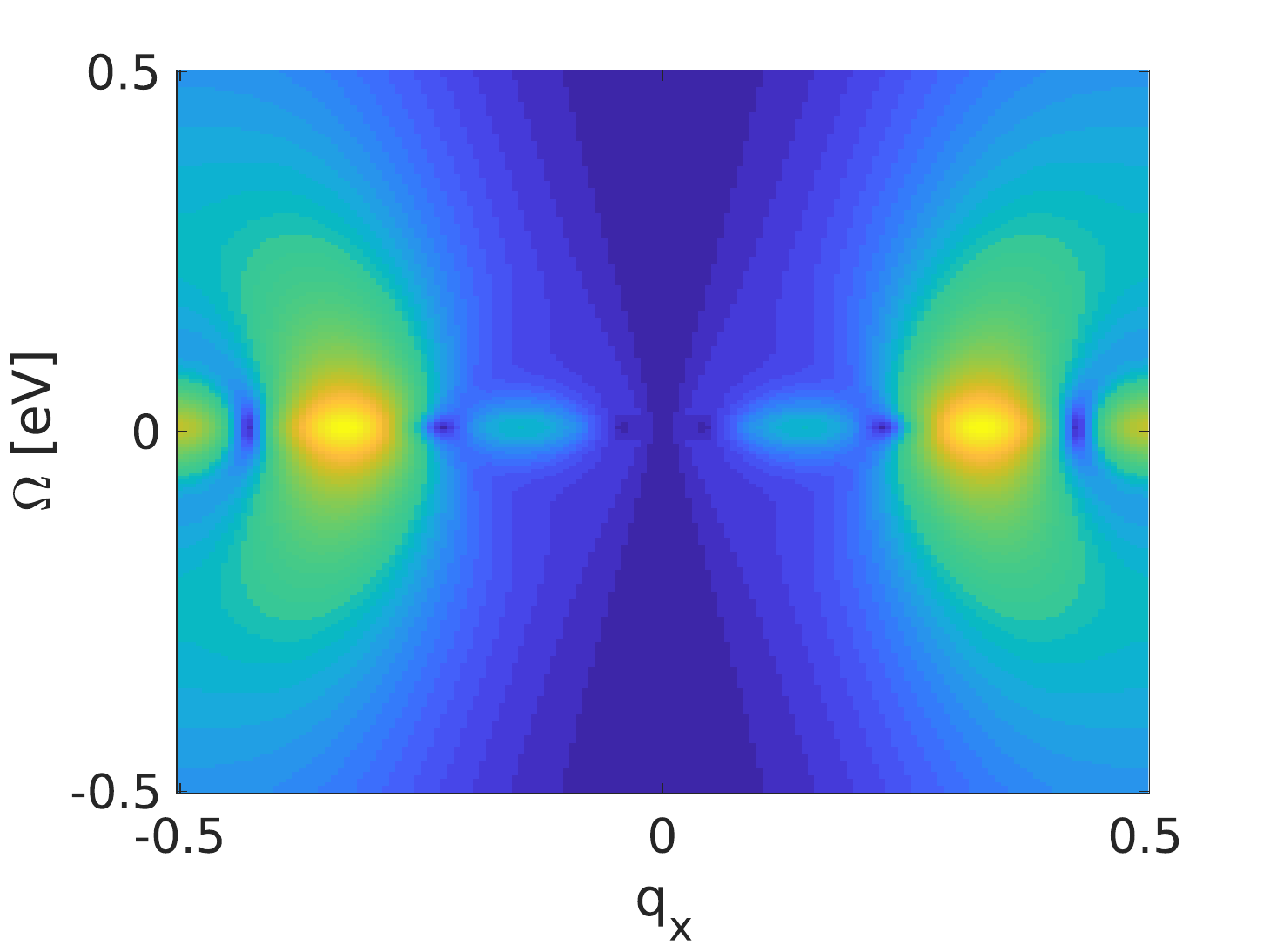} &\includegraphics[width=0.32\textwidth]{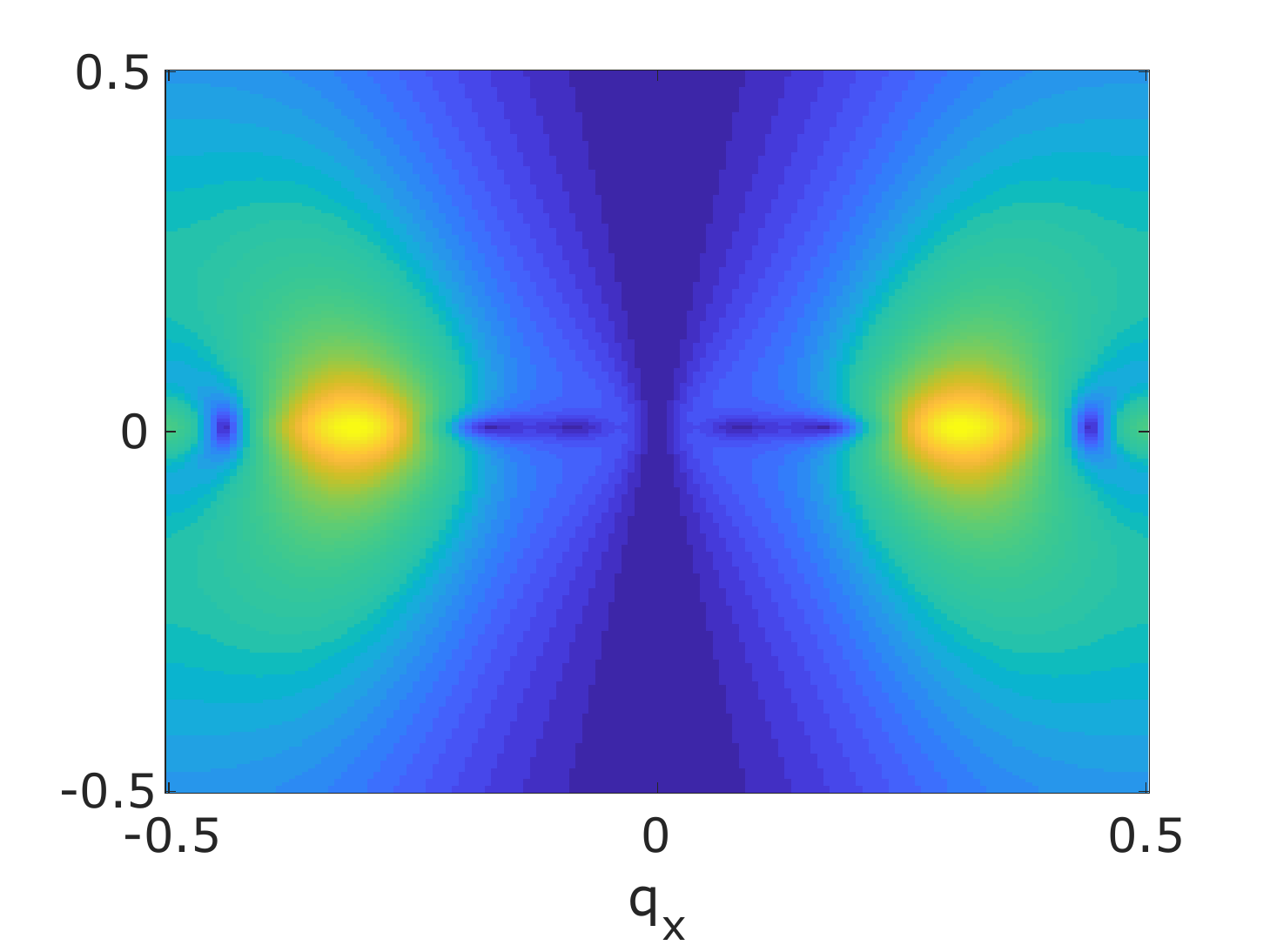} &\includegraphics[width=0.32\textwidth]{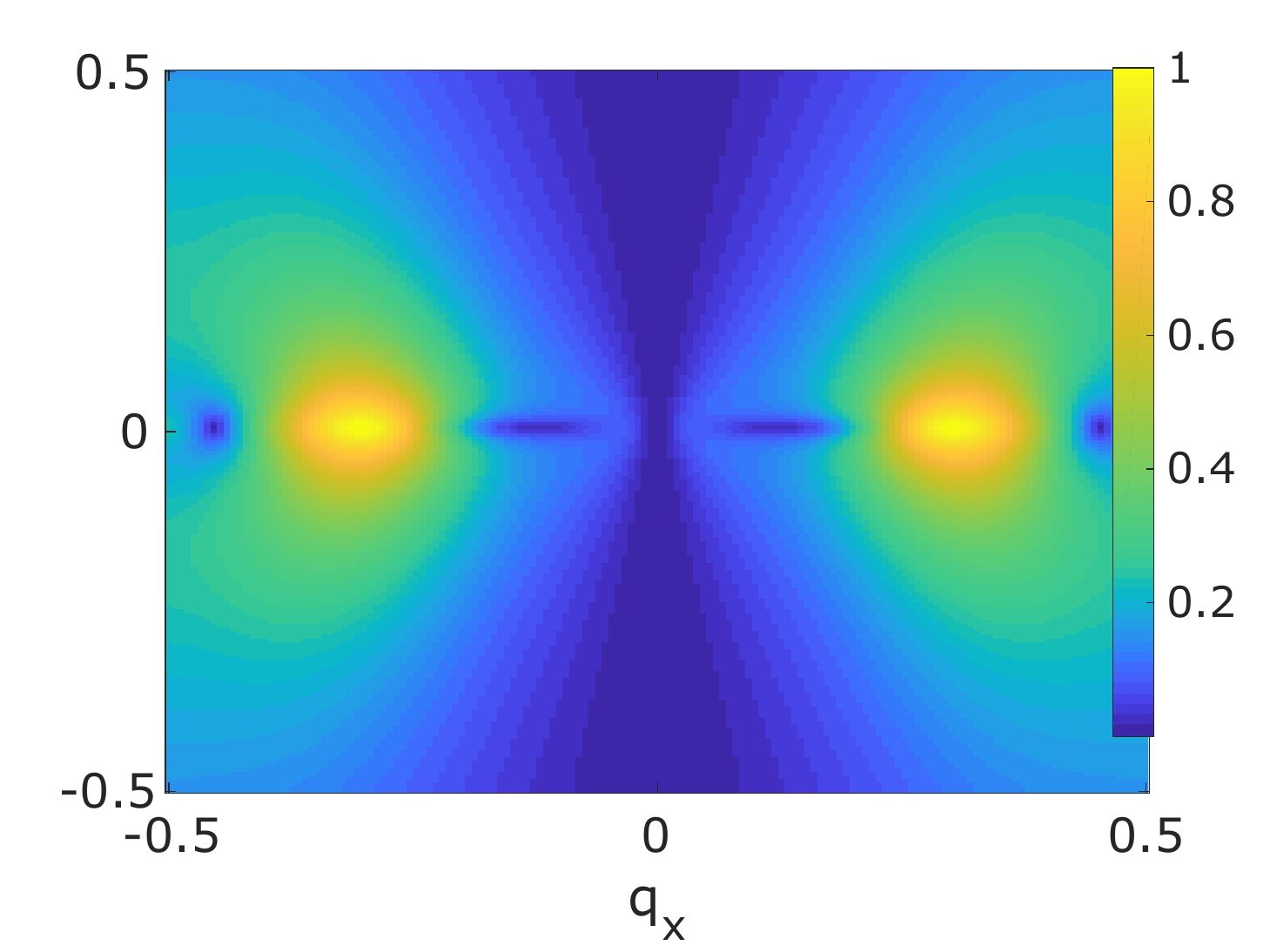}
		\end{tabular}
		\caption{The normalized energy-resolved intensity, Eq.~(\ref{eq:rhoq3}), along the $(q, 0)$ direction. The pronounced quasi-elastic peak ($\Omega \approx 0$ and $q \approx 0.3$ rlu) is substituted by a weaker dispersive signal which shifts towards larger $q$ as the energy is increased.}
		\label{fig:energy}
	\end{figure}

	RIXS experiments involve an exchange of energy between the incoming photons and the electrons. Its intensity has a non-trivial energy dependence, whose microscopic description has been studied in detail (see Refs.~\cite{ament2011resonant, abbamonte2012resonant} for a review). In Ref.~\cite{dentelski2020minimal} we proposed a phenomenological description of the energy dependence in RIXS experiments by evaluating Eq.~(\ref{eq:rhoq3}) at $\Omega=E$ \cite{f1}, see the  lowest panel of Fig.~\ref{Main} where we plot $|\chi(q,\Omega)|$ at $\Omega=0.2$eV. These plots closely resemble the experimental observations at large energy scales, where the peaks at $(\pm q,0)$ and $(0,\pm q)$ become less pronounced. In addition, the ring-like shape becomes wider than the elastic case, indicating a small but noticeable dispersive behavior. This effect can be more clearly seen in Fig.~\ref{fig:energy} where we show the energy dependence of the signal for a one dimensional cut along the $(q, 0)$ direction. As the energy increases, the strong quasi-elastic peak centered around $\Omega\approx 0$ and $q_x \approx 0.3 $ rlu is substituted by a weaker dispersive signal which shifts towards larger $q$.

	Our numerical calculations show that, in contrast to the interpretation provided in Ref.~\cite{boschini2021dynamic}, the experimental observations are consistent with the known Fermi surface of Bi2212, provided that PDW, rather than CDW, oscillations are considered. Our approach calls for a further investigation of this modulation with better energy resolution and in the presence of magnetic field, to refine our understating of the precise nature of these modulations.

	{\it Acknowledgments}. This work is supported by the Israel Science Foundation Grants No. 967/19, No. 151/19 and No. 154/19.


\begin{thebibliography}{30}
\bibitem{f1}
		\BibitemOpen
		\bibfield{footnote} {The Matlab code used to produced the elastic and energy-resolved maps can be found here XXXX}
		\BibitemShut{NoStop}%
\end{thebibliography}
\end{document}